# Further Evidence for Looplike Fine Structure inside "Unipolar" Active Region Plages

Y.-M. Wang, I. Ugarte-Urra, and J. W. Reep
Space Science Division, Naval Research Laboratory, Washington, DC 20375, USA; yi.wang@nrl.navy.mil, ignacio.ugarte-urra@nrl.navy.mil, jeffrey.reep@nrl.navy.mil

## Abstract

Earlier studies using extreme-ultraviolet images and line-of-sight magnetograms from the *Solar Dynamics Observatory* (*SDO*) have suggested that active region (AR) plages and strong network concentrations often have small, looplike features embedded within them, even though no minority-polarity flux is visible in the corresponding magnetograms. Because of the unexpected nature of these findings, we have searched the *SDO* database for examples of inverted-Y structures rooted inside "unipolar" plages, with such jetlike structures being interpreted as evidence for magnetic reconnection between small bipoles and the dominant-polarity field. Several illustrative cases are presented from the period of 2013–2015, all of which are associated with transient outflows from AR "moss." The triangular or dome-shaped bases have horizontal dimensions of ∼2–4 Mm, corresponding to ∼1–3 granular diameters. We also note that the spongy-textured Fe IX 17.1 nm moss is not confined to plages, but may extend into regions where the photospheric field is relatively weak or even has mixed polarity. We again find a tendency for bright coronal loops seen in the 17.1, 19.3, and 21.1 nm passbands to show looplike fine structure and compact brightenings at their footpoints. These observations provide further confirmation that present-day magnetograms are significantly underrepresenting the amount of minority-polarity flux inside AR plages and again suggest that footpoint reconnection and small-scale flux cancellation may play a major role in coronal heating, both inside and outside ARs.

*Unified Astronomy Thesaurus concepts:* Solar coronal heating (1989); Solar magnetic fields (1503); Solar extreme ultraviolet emission (1493); Solar active regions (1974); Solar coronal loops (1485); Solar corona (1483)

*Supporting material:* animations

## 1. Introduction

In a now widely favored model for the heating of coronal loops, random convective motions lead to the formation of current sheets in the overlying magnetic field, where the energy is then dissipated locally (see, e.g., Parker 1988; Galsgaard & Nordlund 1996; Dahlburg et al. 2016). In an alternative (but not necessarily mutually exclusive) scenario, the coronal loops reconnect with small-scale fields at their footpoints, with the dissipated energy being transmitted to greater heights via chromospheric evaporation and magnetohyrodynamic waves (see, e.g., Priest et al. 2018; Syntelis et al. 2019). A list of arguments in favor of the latter view has been given by Aschwanden et al. (2007).

In the case of the active region (AR) corona, a strong argument against footpoint heating is that the plage areas in which the AR loops are embedded generally contain almost no minority-polarity flux, according to even high-resolution magnetograms. Observations recorded by the *Transition Region and Coronal Explorer* in the 17.1 nm passband have shown that plages often have a reticulated appearance, referred to as "moss." It is now widely accepted that AR moss represents the downward-conductively heated, upper transition region of the overlying hot coronal loops (Berger et al. 1999; Fletcher & De Pontieu 1999; Martens et al. 2000; Vourlidas et al. 2001). Rather surprisingly, however, De Pontieu et al. (1999, 2003) found that the 17.1 nm emission is only weakly correlated with the underlying distribution of Ca II K-line brightness, a proxy for the photospheric field strength (see, e.g., Sheeley et al. 2011; Pevtsov et al. 2016).

Recent studies have raised questions about the reliability of present-day magnetograms and the extent to which "unipolar" regions are actually unipolar. In an analysis of extreme-ultraviolet (EUV) coronal plumes employing data from the Atmospheric Imaging Assembly (AIA) and Helioseismic and Magnetic Imager (HMI) on the *Solar Dynamics Observatory* (*SDO*), Wang et al. (2016) found that strong plume emission often occurred above network concentrations where very little minority-polarity flux was visible in the HMI magnetograms. However, simultaneous AIA images invariably showed small, looplike structures in the cores of the plumes. In a subsequent study focusing on unipolar plage areas, Wang (2016) found large numbers of looplike features embedded within the plages, with horizontal sizes of up to ∼3–5 Mm; some of the structures had an inverted-Y topology. Partial support for these unexpected results comes from ∼0″.1-resolution measurements of an emerging AR made with the *SUNRISE* Imaging Magnetograph Experiment (IMaX), as reported by Chitta et al. (2017). They noted that the footpoints of coronal loops were often located near minority-polarity flux that was invisible or barely visible in the corresponding lower-resolution HMI magnetograms (see also Chitta et al. 2018, 2019). However, the minority-polarity elements detected using IMaX were confined to the edges of the strong plage areas, whereas the small looplike features identified in Wang et al. (2016) and Wang (2016) were located well inside the supposedly unipolar plages and network concentrations, as well as at their peripheries.

The *SDO* study of Wang (2016) was limited to three ARs, while the balloon-borne observations of Chitta et al. (2017) were recorded over only 17 minutes. Because of the surprising

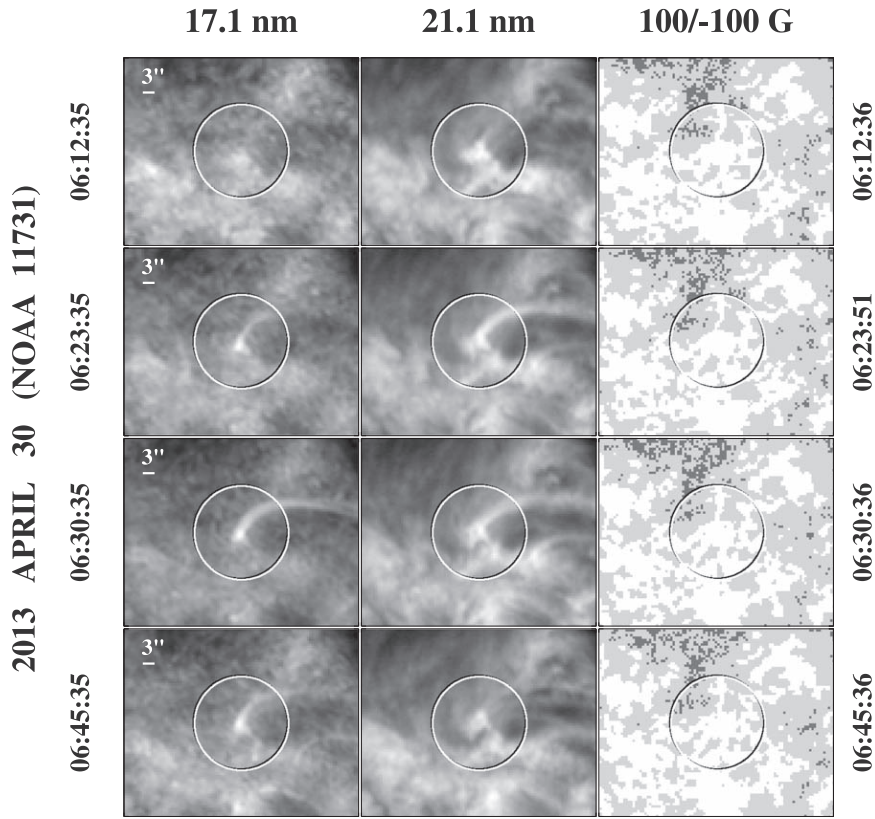

**Figure 1.** Evolution of an inverted-Y structure embedded in unipolar moss/plage in NOAA 11731, 2013 April 30. The circled area has radius 12″ (1″ ≃ 0.7 Mm) and is centered at (−194″, 276″) (06:30 UTC) relative to disk center. North is up and west is to the right. Left column: AIA Fe IX 17.1 nm images recorded at 06:12:35, 06:23:35, 06:30:35, and 06:45:35 UTC. Middle column: simultaneous Fe XIV 21.1 nm images. Right column: HMI line-of-sight magnetograms recorded at 06:12:36, 06:23:51, 06:30:36, and 06:45:36 UTC. Here and in subsequent figures, the grayscale coding used in the magnetograms is as follows: white ($B_{los} > 100$ G); light gray ($0$ G $< B_{los} < 100$ G); dark gray ($-100$ G $< B_{los} < 0$ G); black ($B_{los} < -100$ G). The inverted-Y structure, here seen clearly in both Fe IX ($T \sim 0.7$ MK) and Fe XIV ($T \sim 2.0$ MK), is anchored in strong positive-polarity flux. Its cusp-shaped base has a width and height on the order of 6″ (∼4 Mm). The outflow is first observed in Fe XIV before becoming visible in Fe IX, suggesting that the plasma is initially heated to high temperatures but progressively cools. The available animation presents time sequences, running from 06:00 to 06:59 UTC, of (left to right) the 21.1 nm and 17.1 nm images; the 17.1 nm images with contours of $B_{los}$ superimposed, where the colored lines represent the boundaries between the four $B_{los}$ ranges specified above; and the HMI magnetograms with the same grayscale coding as above.

(An animation of this figure is available.)

nature of some of these recent findings, we have searched the AIA/HMI database for further examples of looplike fine structure inside AR plages. The ARs in our sample were selected randomly from the period of 2013–2016, except for the requirement that the footpoint areas should be clearly visible in Fe IX 17.1 nm, which effectively excluded newly emerged ARs (where the moss tends to be obscured by the overlying bright loops) and those located very close to the solar limb.

## 2. Examples of Inverted-Y Structures inside "Unipolar" AR Plages

The AIA instrument records full-disk images in seven EUV and three UV channels, with 0.″6 pixels and 12 s cadence. HMI provides longitudinal magnetograms with 0.″5 pixels, obtained every 45 s and having a noise level of order 10 G. To search for looplike fine structure inside AR plages, we employ images taken in three passbands: 17.1 nm, dominated by Fe IX ($T \sim 0.7$–0.8 MK), 19.3 nm, dominated by Fe XII ($T \sim 1.5$–1.6 MK), and 21.1 nm, dominated by Fe XIV ($T \sim 1.9$–2.0 MK). The AIA and HMI images were coaligned using the IDL procedures *read_sdo* and *aia_prep* from the SolarSoft library.

In this section, we describe examples of transient outflows from 17.1 nm moss embedded in strong unipolar flux, where the outflows are associated with inverted-Y structures and/or overlie small, loop-shaped features. Such jetlike structures may be seen in almost any AR with visible moss; we have selected some of the clearest cases from a small, arbitrary sample of ARs.

### 2.1. NOAA 11731: 2013 April 30

Figure 1 shows an example of a transient loop structure that is rooted in unipolar AR plage and has a characteristic lifetime on the order of an hour (see also the accompanying animation). The sequence of Fe IX 17.1 nm (Fe XIV 21.1 nm) images in the left (middle) column were recorded between 06:12 and 06:45 UTC on 2013 April 30; the underlying photospheric field is displayed in the right column. Throughout this paper, the grayscale coding for the HMI magnetograms is as follows: white ($B_{los} > 100$ G);

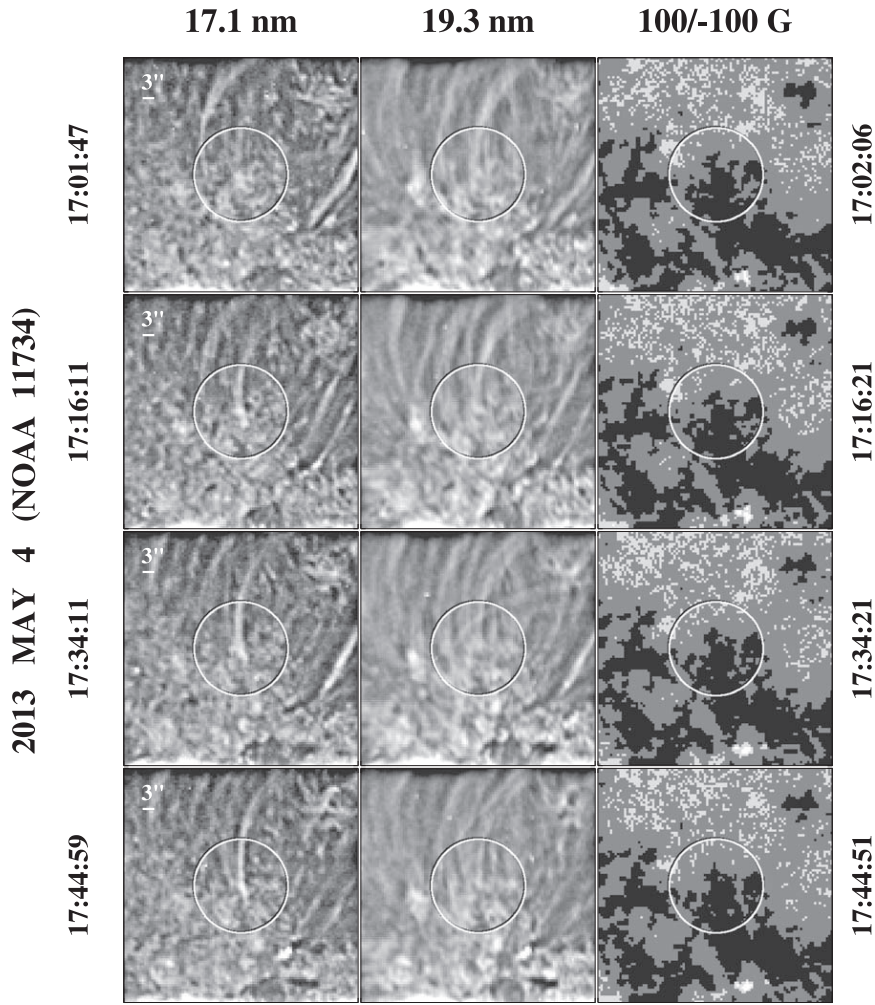

**Figure 2.** Inverted-Y structure rooted in unipolar moss/plage inside NOAA 11734, 2013 May 4. The circled area has radius 12″ and is centered at (−215″, −206″) (17:34 UTC). Left column: sharpened Fe IX 17.1 nm images recorded at 17:01:47, 17:16:11, 17:34:11, and 17:44:59 UTC. Middle column: corresponding images recorded in Fe XII 19.3 nm ($T \sim 1.6$ MK) at 17:01:42, 17:16:06, 17:34:06, and 17:44:54 UTC; again, an unsharp mask has been applied. Right column: HMI magnetograms taken at 17:02:06, 17:16:21, 17:34:21, and 17:44:51 UTC, with the grayscale saturated at ±100 G as in Figure 1. The inverted-Y structure is most clearly seen in the Fe IX images, where it appears in isolation; the Fe XII images show a collection of relatively diffuse loops with complex fine structure around their footpoints.

light gray (0 G < $B_{los}$ < 100 G); dark gray (−100 G < $B_{los}$ < 0 G); black ($B_{los}$ < −100 G).

Initially, the loop is observed as a progressively strengthening outflow from a brightening in the underlying moss, with the outflowing material being more visible in Fe XIV than in Fe IX. Subsequently, the 21.1 nm emission weakens relative to the lower-temperature 17.1 nm emission and is the first to fade. The loop has an inverted-Y topology, with a bright, triangular base suggestive of the "fan" or "dome" of a coronal jet. Located in the positive-polarity plage with field strengths exceeding 100 G, the base has horizontal and vertical dimensions of roughly 6″ (∼4 Mm), corresponding to ∼3 granular diameters. The cusp-like apex appears to rise with time, as might be expected if reconnection or flux exchange between two loop systems is occurring at a null point.

### 2.2. NOAA 11734: 2013 May 4

The Fe IX 17.1 nm images in the leftmost column of Figure 2 show the evolution of another inverted-Y loop, as observed inside NOAA 11734 on 2013 May 4; an unsharp mask has been applied to bring out the fine structure. The base is rooted in negative unipolar plage and appears to consist of one or more small, looplike features. Although not displayed here, subsequent images show the overlying stalk fading after ∼18:00 UTC, leaving a compact brightening at its footpoint.

The middle column of Figure 2 shows the evolution of the same region as it appears in sharpened Fe XII 19.3 nm images. The coronal loops are more diffuse and occupy a wider area of moss in Fe XII ($T \sim 2.0$ MK) than in Fe IX ($T \sim 0.7$ MK). Again, however, small, looplike features and compact brightenings are seen at the footpoints of the large-scale loops. The solitary inverted-Y structure appearing in Fe IX is no longer as distinct in Fe XII, where it is surrounded by similar structures. As in the example of Figure 1, the outflow becomes visible earlier and fades earlier in the higher-temperature passband. The coronal emission becomes even more diffuse and widespread in Fe XIV 21.1 nm images (not displayed here).

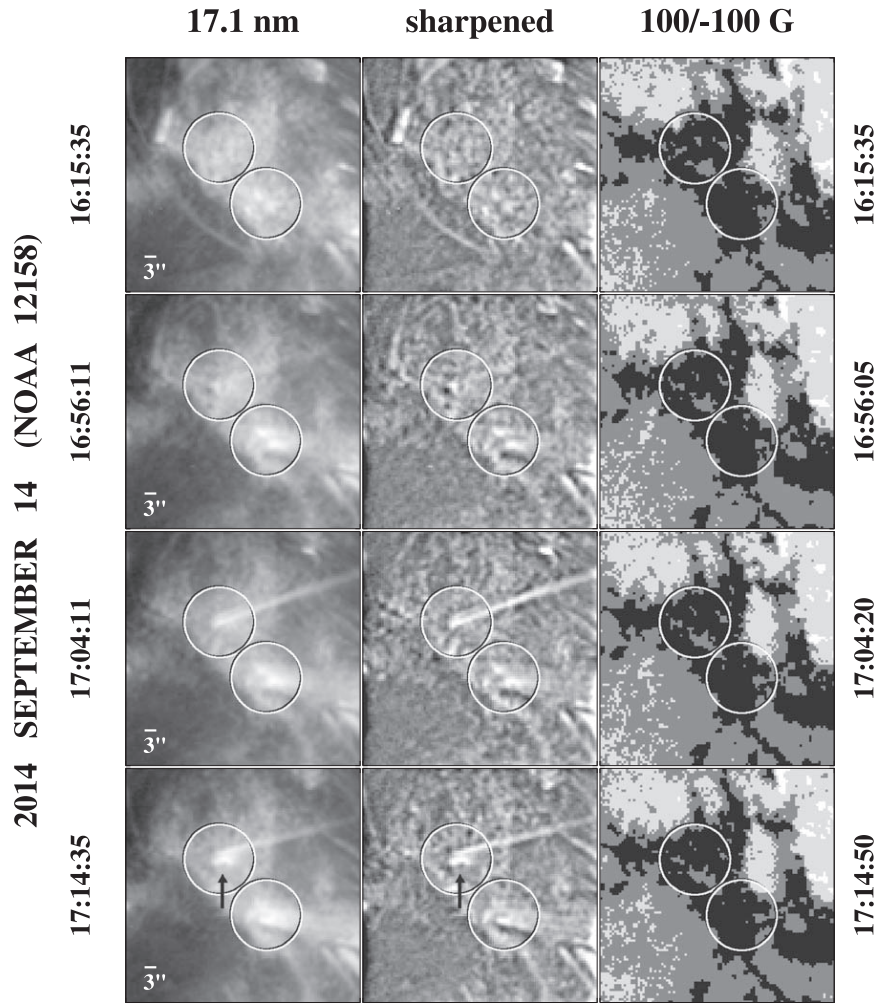

**Figure 3.** Transient loop outflows observed inside NOAA 12158 during 2014 September 14. The circled areas have radii of 9″ and are centered at (634″, 149″) and (646″, 135″) (17:14 UTC). Left column: Fe IX 17.1 nm images recorded at 16:15:35, 16:56:11, 17:04:11, and 17:14:35 UTC. Middle column: sharpened versions of the same images. Right column: HMI longitudinal magnetograms recorded at 16:15:35, 16:56:05, 17:04:20, and 17:14:50 UTC and saturated at ±100 G. In the 17:14 UTC panels, the arrow points to a clearly defined loop situated next to the footpoint of the long 17.1 nm stalk. This small, bright loop has a horizontal extent of ∼4″ and is rooted in purely unipolar plage, according to the corresponding HMI magnetogram. The available animation presents time sequences, running from 16:40 to 17:29 UTC, of (left to right) the 17.1 nm images; the 17.1 nm images with contours of $B_{los}$ superimposed; and the HMI magnetograms. The gray scale for the magnetograms and the colored contour lines are as indicated in Figure 1's caption.

(An animation of this figure is available.)

### 2.3. NOAA 12158: 2014 September 14

The Fe IX 17.1 nm images in the left column of Figure 3 show the formation of three stalk-like structures inside NOAA 12158 on 2014 September 14; in the middle column, an unsharp mask has been applied to the same sequence of images, which were recorded between 16:15 and 17:14 UTC. At 17:14 UTC (bottom panels), a small, bright loop is clearly visible within the circled area just to the northeast of the image center. This loop, located next to the base of the narrow 17.1 nm stalk, has a horizontal extent of ∼4″ and is rooted entirely within the negative-polarity plage. It is not as well defined in 19.3 or 21.1 nm images (not displayed here). A comparison of the images taken at 17:04 and 17:14 UTC suggests that the looplike feature was earlier attached to the stalk but then separated from it (see also the accompanying animation). The separation might be due to the fading of the original stalk/jet and the strengthening of the adjacent outflow. We also note that reconnection between small- and large-scale loop systems would lead to an exchange of footpoints and changes in the intensities and orientations of the reconnecting structures.

### 2.4. NOAA 12221: 2014 December 1

The 17.1 nm images in the left column of Figure 4 show a number of inverted-Y structures inside NOAA 12221 during 2014 December 1; the middle column displays sharpened versions of the same images. At 01:22 UTC, in the southwest quadrant of the circled area, a small loop of width ∼3″ may be seen embedded in the positive-polarity plage at the base of a narrow outflow stream. Another inverted-Y structure is visible to the northeast, whose eastern leg is located above a negative-polarity region. In the image taken at 02:14 UTC, the long, newly formed stalk rooted inside the circle has a topologically complex base that is centered above an area where the unipolar field is relatively weak (<100 G). At 02:37 UTC, the HMI magnetogram shows some minority-polarity flux elements invading this weaker-field corridor, but the shrunken base of the narrow stalk remains embedded in positive-polarity flux.

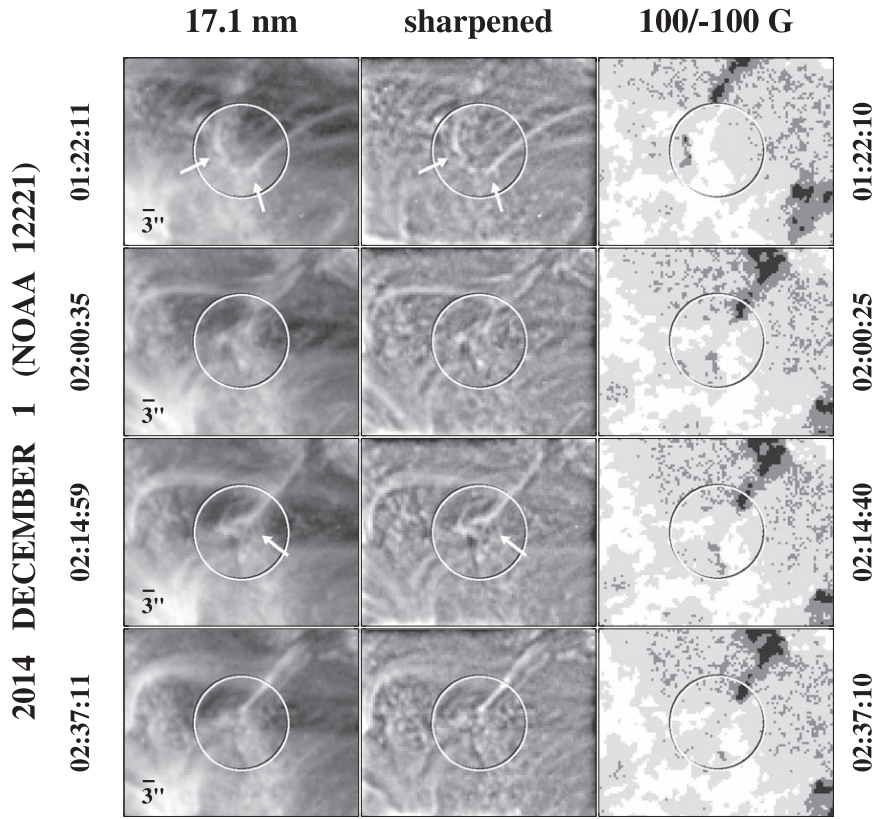

**Figure 4.** Inverted-Y structures observed inside NOAA 12221 on 2014 December 1. Circles have a radius of $12''$ and are centered at $(-324'', 64'')$ (01:22 UTC) and $(-323'', 70'')$ (02:14 UTC). Left column: Fe IX 17.1 nm images recorded at 01:22:11, 02:00:35, 02:14:59, and 02:37:11 UTC. Middle column: sharpened versions of the same images. Right column: HMI magnetograms recorded at 01:22:10, 02:00:25, 02:14:40, and 02:37:10 UTC and saturated at $\pm 100$ G. In the 17.1 nm images taken at 01:22 UTC, the inverted-Y structure toward the eastern edge of the circled area has one leg rooted in or near an area of minority-polarity flux.

### 2.5. NOAA 12232: 2014 December 13–15

The (original and sharpened) Fe IX 17.1 nm images in Figure 5, recorded at 14:06 UTC on 2014 December 13, show a row of three inverted-Y structures inside NOAA 12232. Their bases have horizontal dimensions of $\sim 3''$, close to the typical granular diameter of $\sim 2''$. The southernmost structure overlies an area containing minority-polarity flux, but the two structures to the north are rooted entirely inside positive-polarity plage or strong network. The middle Fe IX structure is not visible in Fe XII 19.3 nm (third panel). It is also noteworthy that the bright 17.1 nm moss underlying these structures extends southward into an area of relatively weak, mixed-polarity flux; this is consistent with the idea that the moss contains small loops (see also Figures 8 and 11 below).

Figure 6 displays 17.1 and 19.3 nm observations of the same AR during 10:31–11:01 UTC on December 15. Toward the top of the circled area, the 17.1 nm image recorded at 10:31 UTC shows a pair of closely spaced stalks with a suggestion of cusp-like features at their bases. By 10:51 UTC, these narrow inverted-Y structures have largely faded away. In the meantime, a brightening near the center of the circled area increases in size and intensity and develops into a 17.1 nm stalk with a large, bulbous base having horizontal dimensions of $\sim 6''$ or $\sim 4$ Mm (see the bottom left image recorded at 11:01 UTC). As observed in the 19.3 nm passband, this structure was already present at 10:31 UTC, but continually brightens over the next 30 minutes. The dome-shaped base (which may actually consist of an arcade of small loops) is entirely embedded within the positive-polarity plage. Near the bottom of the circled area, the sharpened 19.3 nm images also show a loop-shaped feature that appears to straddle the boundary between the positive- and negative-polarity regions.

### 2.6. NOAA 12436: 2015 October 23

The sequence of Fe IX 17.1 nm images in the left column of Figure 7 show the formation of an inverted-Y structure inside NOAA 12436 on 2015 October 23. The dome-like base is rooted in positive-polarity plage and has an angular size of $\sim 6''$. Initially, at 01:10 UTC, a brightening in the underlying moss is observed, which evolves into a small, curved feature at 01:14 UTC. The narrow stalk seen at 01:22 UTC greatly widens by 10:27 UTC. The corresponding Fe XIV 21.1 nm images in the middle column of Figure 7 show a broad, diffuse outflow from the same area throughout this period, with the base structure being topologically similar to that observed in the lower-temperature bandpass. (In the 17.1 nm image taken at 01:10 UTC, a faint, narrow "jet" with a cusp-shaped base is visible just to the west of the moss brightening.)

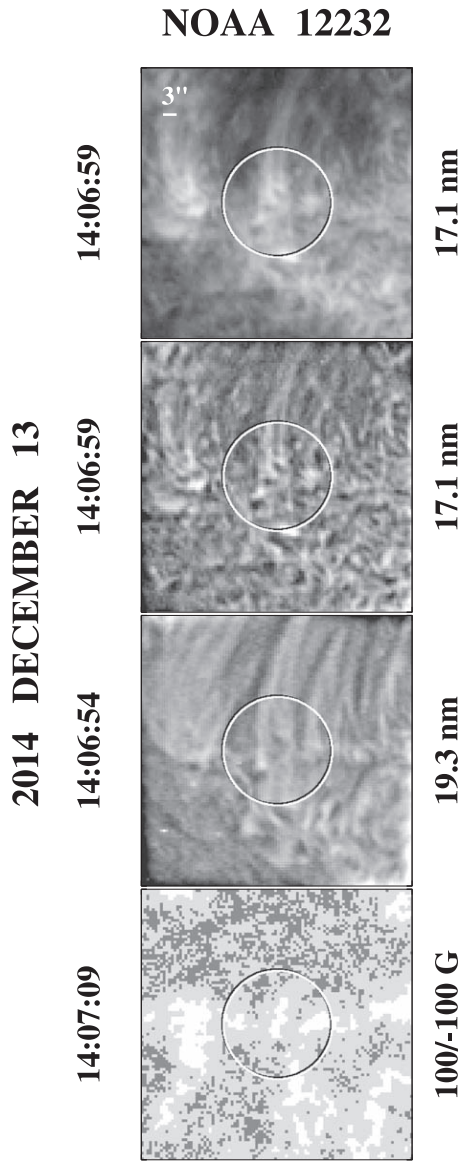

**Figure 5.** Inverted-Y structures observed in NOAA 12232 on 2014 December 13. The circled area has radius $12''$ and is centered at $(-160'', 223'')$. Top panel: Fe IX 17.1 nm image recorded at 14:06:59 UTC. Second panel: sharpened version of the 17.1 nm image. Third panel: sharpened Fe XII 19.3 nm image recorded at 14:06:54 UTC. Bottom panel: HMI magnetogram taken at 14:07:09 UTC and saturated at $\pm 100$ G. Within the circled area, the 17.1 nm images show a row of three inverted-Y structures with base widths of order $3''$; the southernmost structure overlies a small pocket of minority/negative-polarity flux, but the other two structures are rooted in positive-polarity plage/network. The 19.3 nm image does not show the middle structure, although a number of other loops with bulbous bases are seen in and around the circled area. Comparing the top and bottom panels, we note that the bright 17.1 nm moss extends south- and southeastward from the center of the circled area into a region of weak, mixed-polarity flux.

### 2.7. NOAA 12470: 2015 December 19

The Fe IX and Fe XII images in Figure 8 show narrow, transient outflows from a mossy region inside NOAA 12470 during 2015 December 19. At 05:34 UTC, a compact, short-lived brightening (arrowed) is observed above the base of the outflow/jet; from the HMI magnetogram, note the presence of small minority-polarity flux elements just northward of the base. The location of the compact brightening is roughly coincident with that of the cusp-like feature indicated by the arrow in the 05:36 UTC frame. At 05:45 UTC, a small, bright loop having a width of $\sim 3''$ is clearly visible under the 17.1 nm jet. The accompanying animation shows weak negative-polarity elements just southward of the base region impinging on the positive-polarity plage and disappearing, as a result of either flux cancellation or inadequate spatial resolution/instrument sensitivity. As in the example of Figure 5, the 17.1 nm images in Figure 8 show bright moss extending even into the regions of mixed polarity to the east and southeast of the circled area (see also Figure 11 below).

### 3. Looplike Fine Structure at the Base of Bright Coronal Loops

In the preceding section, we displayed examples of transient outflows occurring above loop- or cusp-shaped features rooted inside supposedly unipolar plage areas. This arbitrary sample from the AIA/HMI image archive provides further support for the assertion that AR plages contain significant amounts of minority-polarity flux that does not appear in the HMI magnetograms. The following examples pertain more generally to the relationship between the looplike fine structure observed in AR moss and the heating of the overlying large-scale loops.

#### 3.1. NOAA 11734: 2013 May 5

The 17.1 and 19.3 nm images in Figure 9 focus on a bundle of long coronal loops embedded in a unipolar plage area within NOAA 11734. The observations were taken between 00:07 and 00:45 UTC on 2013 May 5. In the sharpened 17.1 nm images, small, curved features and compact brightenings may be seen at the footpoints of the loops. This fine-scale ($\sim 3''$–$4''$) structure appears to change continually over the 38 minute interval, as does the detailed spatial and intensity distribution of the overlying loops. The Fe XII loops shown in the third column of Figure 9 are wider and more diffuse than their Fe IX counterparts, but likewise exhibit time-varying, topologically complex structure at their bases.

#### 3.2. NOAA 12443: 2015 November 5

As another example, the left column of Figure 10 shows a collection of Fe IX 17.1 nm loops at four different times during late 2015 November 5; the middle column displays the corresponding Fe XII 19.3 nm images. The loops are located in an area of unipolar plage within NOAA 12443. Again, a continually evolving configuration of compact brightenings and small, looplike features, with scale sizes of $\sim 3''$–$6''$, may be seen at the footpoints of the 17.1 nm loops. These fine-structure changes are accompanied by variations in the overlying coronal emission; for example, the arrowed footpoint feature in the bottom left panel of Figure 10 underlies a newly brightened 17.1 nm stalk (see the accompanying animation, which also shows downflows occurring as Fe IX loops fade). As observed at higher temperatures, the coronal emission is more diffuse and ubiquitous, but complexly structured brightenings may still be seen at the footpoints of the Fe XII loops.

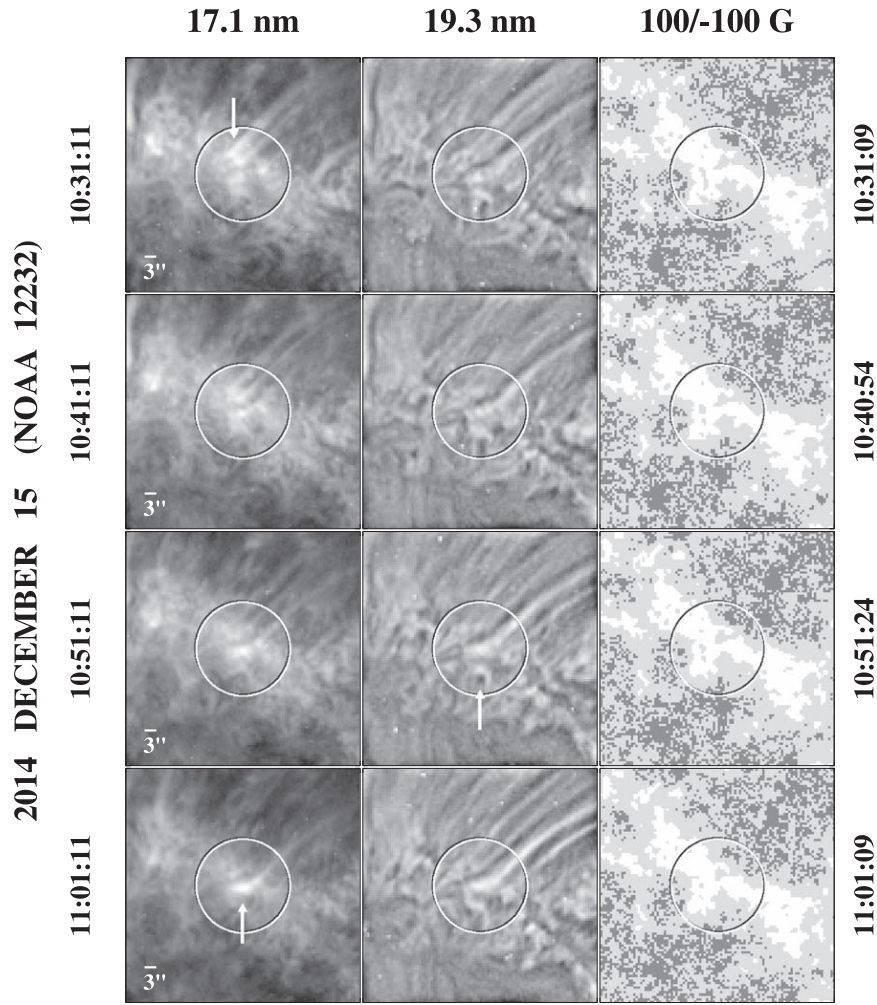

**Figure 6.** Inverted-Y structures inside an area of moss/plage in NOAA 12232, 2014 December 15. Circles have radii 12″ and are centered at (264″, 218″) (11:01 UTC). Left column: Fe IX 17.1 nm images recorded at 10:31:11, 10:41:11, 10:51:11, and 11:01:11 UTC. Middle column: sharpened Fe XII 19.3 nm images recorded at 10:31:06, 10:41:06, 10:51:06, and 11:01:06 UTC. Right column: HMI magnetograms taken at 10:31:09, 10:40:54, 10:51:24, and 11:01:09 UTC and saturated at ±100 G. Toward the top of the circled area, the 17.1 nm images show a pair of narrow stalks (also visible in the 19.3 nm passband) with cusp-like bases, which fade between 10:31 and 11:01 UTC. Meanwhile, near the center of the circled area, another jetlike structure forms, whose bright, dome-shaped base has a width of ∼6″ (∼4 Mm). This structure (which may contain separate components that merge in the line of sight) is also prominent in the 19.3 nm images, where it is already visible at 10:31 UTC. A variety of small, looplike features may be seen within the 19.3 nm moss, some of which appear to have one leg rooted in the negative-polarity flux at the edge of the positive-polarity plage (see, e.g., the arrowed feature at 10:51 UTC).

## 4. Relationship Between Moss and the Photospheric Field Strength

An inspection of the previous figures suggests that, while areas of bright 17.1 and 19.3 nm moss frequently overlie a strong (>100 G) photospheric field, they may also occur where the underlying field is relatively weak. Conversely, areas of strong unipolar flux sometimes appear dark in the 17.1 nm bandpass; this is usually due to the presence of horizontal fibril structure in the vicinity of polarity inversions.

Figure 11 focuses on an area of moss observed inside NOAA 12232 on 2015 December 15 (this region is located near the top left corner of the images in Figure 6). The bright 17.1 nm moss within the circle is seen to overlie a patch of weak, mixed-polarity field. Morphologically, it is virtually indistinguishable from the adjoining moss that extends over the positive-polarity plage to the southwest, where the field strengths are in excess of 100 G. This similarity in appearance suggests that moss is not simply a tracer of the footpoints of hot coronal loops embedded in strong unipolar flux. (Figures 5 and 8 provide further examples of bright moss that extends into areas of mixed polarity.)

As remarked in Wang (2016), moss-like regions are also found outside ARs. Figure 12 shows the 17.1, 19.3, and 21.1 nm emission associated with a chain of negative-polarity network elements located just to the north of NOAA 12321, as observed at 14:29 UTC on 2015 April 18. In the sharpened 17.1 nm image, inverted-Y structures and small looplike features may be seen within a "mossy" region that is oriented roughly along the axis of the supergranular network elements. In this case, the presence of small loops is hardly surprising,

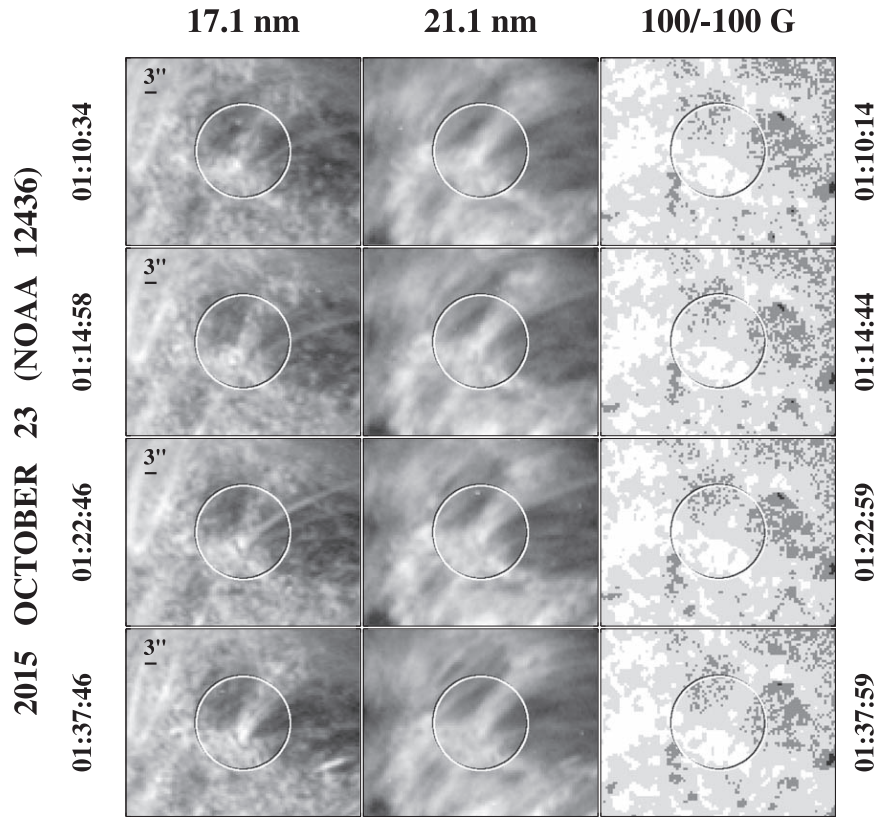

**Figure 7.** Formation of a jetlike structure in NOAA 12436, 2015 October 23. The circled area has a radius of 12″ and is centered at (−149″, 100″) (01:22 UTC). Left column: Fe IX 17.1 nm images taken at 01:10:34, 01:14:58, 01:22:46, and 01:37:46 UTC. Middle column: simultaneous Fe XIV 21.1 nm images. Right column: HMI magnetograms taken at 01:10:14, 01:14:44, 01:22:59, and 01:37:59 UTC. At 01:22 UTC, the 17.1 nm image shows a long, narrow "jet" with a dome-shaped base, whose looplike nature is also suggested by the 17.1 and 21.1 nm images recorded at 01:14 UTC. The base has an angular size of ∼6″ and is embedded in positive-polarity plage. The outflow observed in the 17.1 nm passband widens considerably between 01:22 and 01:37 UTC. The 19.3 nm images show diffuse emission over the entire base area throughout the interval 01:10–01:37 UTC.

given that the underlying network is interspersed with minority-polarity flux elements.

Finally, Figure 13 shows an inverted-Y structure embedded in an area of moss inside NOAA 12335, as observed at 06:14 UTC on 2015 May 5. The base of the structure is rooted in strong negative-polarity plage, but the bright moss occupies a much wider area, extending through the weaker-field region to the south. The dark areas in the 17.1 nm image in the top panel represent fibrilar and filament material located near large- and small-scale polarity inversion lines.

These examples suggest that a strong underlying photospheric field is neither a necessary nor a sufficient condition for the presence of bright moss.

## 5. Summary and Discussion

We have presented illustrative examples of inverted-Y structures embedded inside Fe IX 17.1 nm moss, where the underlying photospheric field was purely unipolar according to simultaneous HMI magnetograms. (Additional examples may be seen in Figures 2, 4, and 5 of Wang 2016.) The triangular or dome-shaped bases of these structures had horizontal dimensions of ∼3″–6″ or ∼2–4 Mm. In the standard topological interpretation, the base contains minority-polarity flux, with a null point located at its apex. The lifetimes of the looplike footpoint features ranged from less than a minute to the order of an hour, with the smaller features generally being shorter-lived than the larger ones.

As an alternative possibility that does not require the presence of minority-polarity flux, the triangular structures might represent canopies overlying individual granular cells, with thermal pressure dominating above the cell centers and excluding the unipolar flux fanning out from the granular lanes. However, the base widths of some of our inverted-Y structures are well in excess of a granular diameter (∼2″). Moreover, the observed cusps are generally located at heights above ∼2 Mm, where the plasma $\beta$ falls well below unity and thermal pressure becomes negligible (see, e.g., Figure 3 in Gary 2001). In addition, in cases such as that displayed in the bottom panels of Figure 3, small loops are sometimes unmistakably present.

We are thus led to conclude that AR plages may contain minority-polarity flux that does not appear in HMI longitudinal magnetograms. That the footpoints of the looplike features are often separated by as many as ∼5–10 pixels suggests that the failure to detect the minority polarity is not simply a matter of spatial resolution, but may also be related to instrument sensitivity. We conjecture that pixels containing strong flux of the dominant polarity sometimes "bleed" into

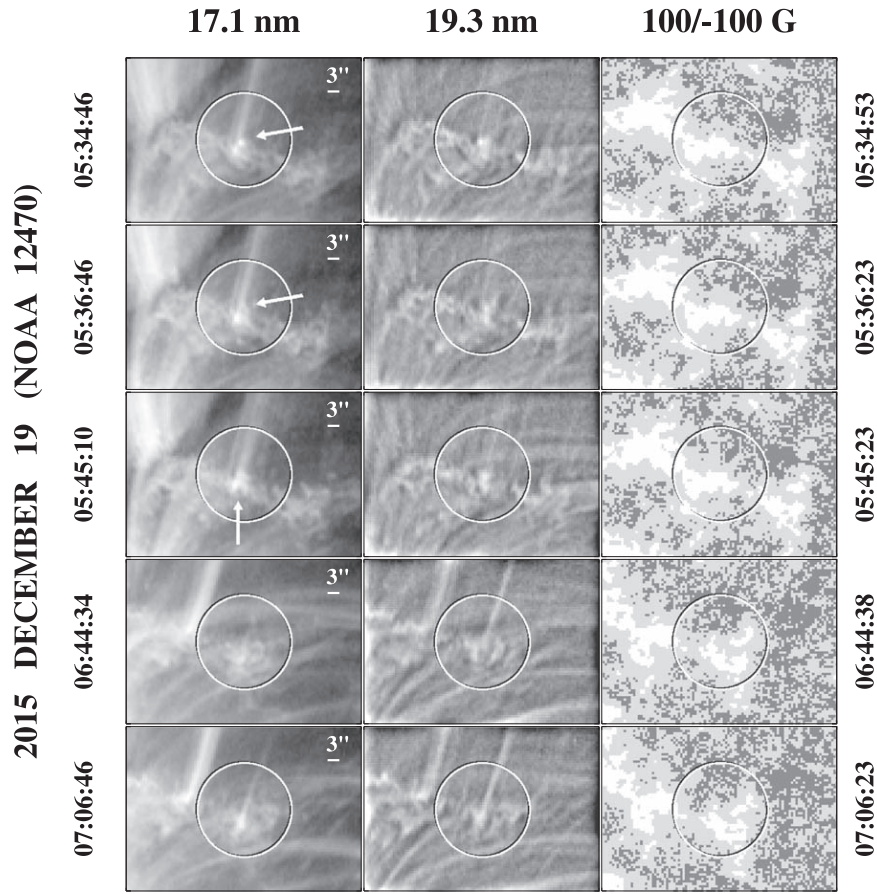

**Figure 8.** Transient outflow events inside NOAA 12470 during 2015 December 19. The circled areas have radii 12″ and are centered at (−101″, 346″) (05:45 UTC) and (−76″, 343″) (07:06 UTC). Left column: Fe IX 17.1 nm images recorded at 05:34:46, 05:36:46, 05:45:10, 06:44:34, and 07:06:46 UTC. Middle column: sharpened Fe XII 19.3 nm images recorded at 05:34:41, 05:36:41, 05:45:05, 06:44:29, and 07:06:41 UTC. Right column: HMI magnetograms recorded at 05:34:53, 05:36:23, 05:45:23, 06:44:38, and 07:06:23 UTC. In the 17.1 nm image taken at 05:34 UTC, a compact brightening is seen above the base of the stalk/jet; this short-lived brightening occurs near the location of the cusp-like feature indicated by the arrow in the 05:36 UTC frame. At 05:45 UTC, a small, bright loop is clearly visible under the 17.1 nm jet, with minority-polarity flux elements present just southward of it. Throughout this period, both the 17.1 and the 19.3 nm images show looplike features pervading the moss, which extends even into the regions of mixed polarity to the east and southeast of the circled area. The available animation presents time sequences, running from 05:19 to 05:59 UTC, of (left to right): the 17.1 nm images, the 17.1 nm images with contours of $B_{los}$ superimposed, and the HMI magnetograms. The gray scale for the magnetograms and the colored contour lines are as indicated in Figure 1's caption.

(An animation of this figure is available.)

adjacent pixels containing much weaker flux of the opposite polarity. In the case of the HMI measurements, each 45 s line-of-sight magnetogram is constructed from 72 individual filtergrams taken over a 270 s period; the procedure involves correcting for differential rotation and interpolating in both space and time (see Couvidat et al. 2016). However, while the interpolation may have slightly weakened any minority-polarity signal, the problem is likely to be of a more fundamental nature, involving the original spectral line profiles and their physical interpretation.

Although the looplike features described here have horizontal extents ≳3″, some of the 17.1 nm images in this paper and in Wang (2016) suggest the presence of loops even closer to the ∼1″ resolution limit. From their analyses of 19.3 nm images recorded with ∼0.″3–0.″4 resolution by the rocket-borne High-resolution Coronal Imager, Peter et al. (2013) and Barczynski et al. (2017) have found evidence for granular-size coronal loops with footpoint separations of only ∼1 Mm. On this spatial scale, it is hardly surprising that the corresponding HMI magnetograms showed the same polarity at both loop footpoints.

Section 3 focused on collections of large-scale loops that were rooted in unipolar flux and that underwent brightness changes on timescales of minutes to an hour. These intensity variations were accompanied by changes in the fine-scale structure of the loop footpoints, where compact brightenings and small looplike features were observed. We attribute this continually evolving footpoint topology to the effect of reconnection with minority-polarity flux that does not appear in the HMI magnetograms. An alternative interpretation might be that the loop legs are being directly deformed by the underlying granular convection itself, rather than by reconnection processes driven by this convection or by small-scale flux emergence. However, deformations due to granular jostling would be propagated away by magnetohydrodynamic waves and could not account for the compact brightenings and cusp-shaped features that extend to heights ≳2 Mm, where $\beta \ll 1$.

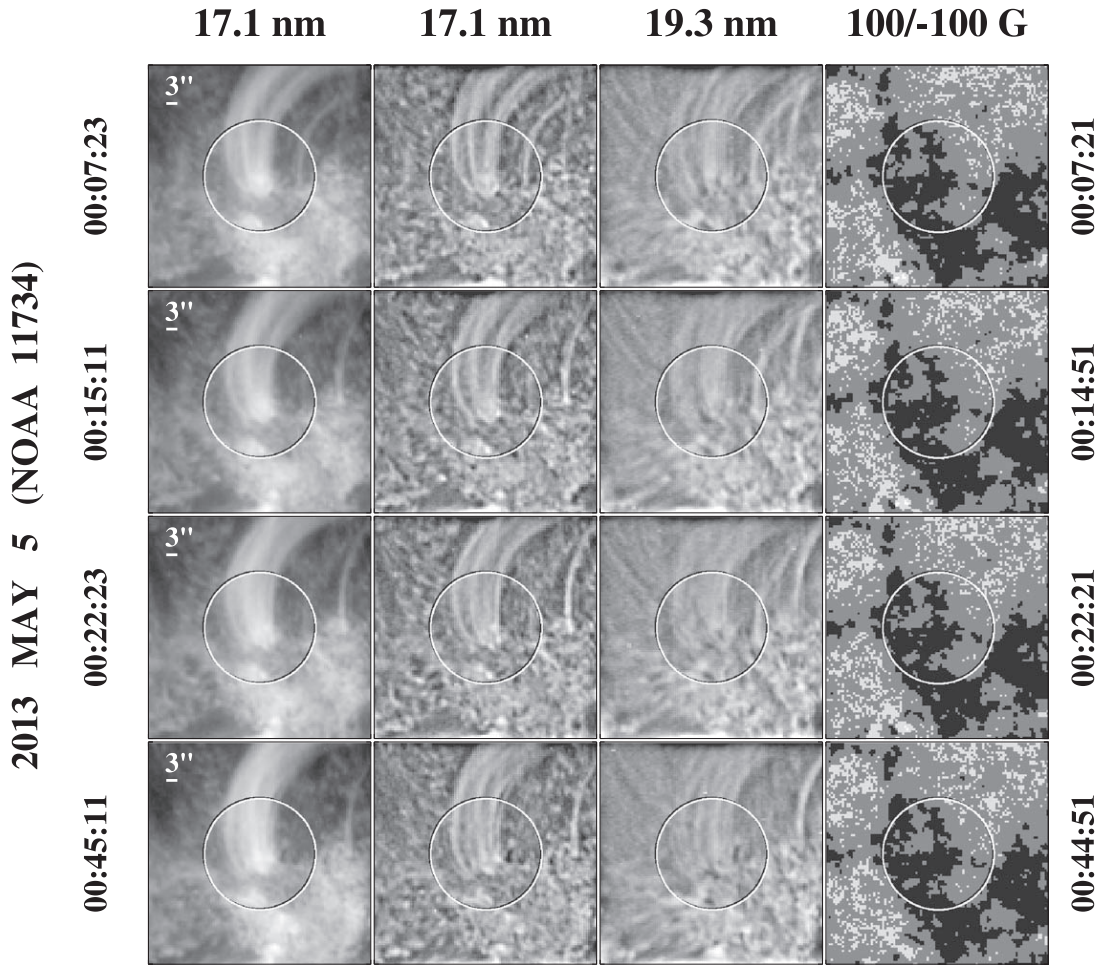

**Figure 9.** Time-varying footpoint structure below a bundle of coronal loops inside NOAA 11734, 2013 May 5. The circled area has a radius of 15″ and is centered at (−176″, −208″) (00:22 UTC). Leftmost column: Fe IX 17.1 nm images taken at 00:07:23, 00:15:11, 00:22:23, and 00:45:11 UTC. Second column: sharpened versions of the same images. Third column: sharpened Fe XII 19.3 nm images taken at 00:07:18, 00:15:06, 00:22:18, and 00:45:06 UTC. Rightmost column: HMI magnetograms recorded at 00:07:21, 00:14:51, 00:22:21, and 00:44:51 UTC and saturated at ±100 G. The curved features and compact brightenings underlying the long 17.1 nm loops have horizontal dimensions typically on the order of 3″–4″. As observed in Fe XII, the bundle of coronal loops is broader and more diffuse than in Fe IX, but the footpoint region continues to exhibit a complex topology.

As mentioned in the Introduction, De Pontieu et al. (1999, 2003) found that AR moss emission is poorly correlated with Ca II K-line brightness, a proxy for the unsigned photospheric flux. We have shown examples of bright moss overlying relatively weak field, including even areas of mixed polarity (see Figures 5, 8, and 11). Such observations provide support for the hypothesis that the moss emission depends not just on the strength of the photospheric field, but also on the presence of minority-polarity flux, which in turn gives rise to the looplike fine structure seen in mossy regions both inside and outside ARs.

In the transient outflow events described in Section 2, we found a general tendency for the outflows to appear in the higher-temperature passbands (21.1 and 19.3 nm) well before becoming visible at 17.1 nm. This behavior may be understood if the evaporating chromospheric material is initially heated to high temperatures, but subsequently cools (see, e.g., Ugarte-Urra et al. 2006; Aschwanden et al. 2007). The sporadic and isolated nature of some of our events suggests that they may have been triggered by underlying flux emergence. It is also well known that lower-temperature loops tend to be more dynamic, with faster cooling rates and shorter lifetimes, than high-temperature loops (see Ugarte-Urra & Warren 2012, and references therein).

We have focused on relatively cool ($T \lesssim 2$ MK) coronal loops because they are more easily identified and tracked than the diffuse, high-temperature loops that occupy most of the moss (Berger et al. 1999; Fletcher & De Pontieu 1999). As suggested by the 17.1 nm images in Figure 1, many areas of brightened moss do not show overlying loops in the lower-temperature passbands. Based on our examples involving "activated" 17.1, 19.3, and 21.1 nm loops, we infer that the moss emission is enhanced where minority-polarity flux undergoes reconnection with the dominant-polarity field. If these footpoint reconnection events heat the overlying loops to temperatures well in excess of ∼2 MK, lower-temperature loops will be observed only if the hot loops have sufficient time to cool before the next reconnection event (see Ugarte-Urra et al. 2009).

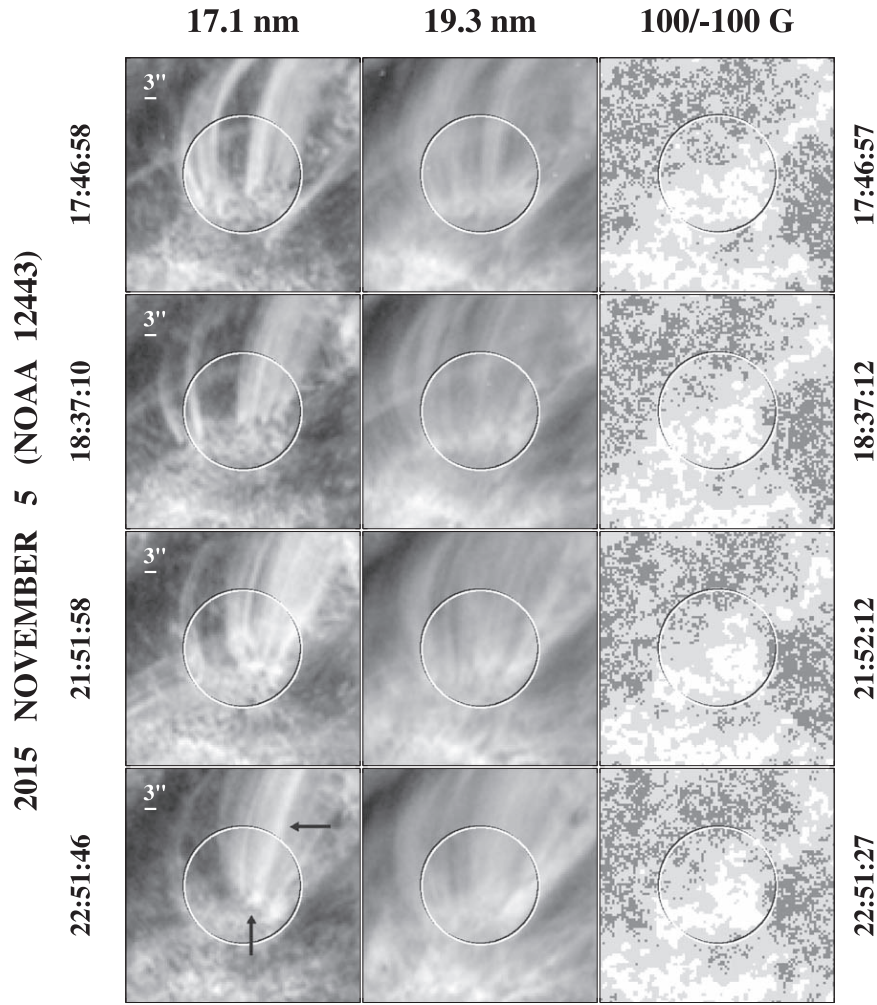

**Figure 10.** Time-varying cluster of coronal loops observed inside NOAA 12443 on 2015 November 5. The circled area has radius 15″ and is centered at (334″, 142″) (21:51 UTC). Left column: Fe IX 17.1 nm images recorded at 17:46:58, 18:37:10, 21:51:58, and 22:51:46 UTC. Middle column: Fe XII 19.3 nm images recorded at 17:46:53, 18:37:05, 21:51:53, and 22:51:41. Right column: HMI magnetograms taken at 17:46:57, 18:37:12, 21:52:12, and 22:51:27 UTC and saturated at ±100 G. The brightness and spatial distribution of the 17.1 nm loops vary continually over the 5 hr period, as does the underlying fine structure, which consists of looplike features and compact brightenings. Arrows in the 17.1 nm image recorded at 22:51 UTC point to a bright outflow stream with a compact brightening at its base. Animations suggest that the fading of the Fe IX loops is often accompanied by downflows. At progressively higher temperatures, the coronal emission becomes increasingly diffuse and widespread; however, the Fe XII loops continue to exhibit complexly structured brightenings at their bases. Most of the footpoint structures are rooted in unipolar plage. The available animation presents time sequences, running from 20:59 to 22:58 UTC, of (left to right): the 17.1 nm images, the 17.1 nm images with contours of $B_{los}$ superimposed, and the HMI magnetograms. The gray scale for the magnetograms and the colored contour lines are as indicated in Figure 1's caption.

(An animation of this figure is available.)

Additional heating mechanisms, such as current sheet formation (field-line braiding) due to footpoint shuffling, may well be operating in coronal loops. However, because the coronal field tends to weaken and become more uniform with increasing height, reconnection near the loop footpoints, where the field is both strong and topologically complex, is likely to play a major role provided that sufficient minority-polarity flux is present. This paper has presented further evidence that the amount of minority-polarity flux inside AR plages has been underestimated.

To investigate further the reasons for the discrepancies between the EUV and magnetic data, it would be worthwhile to compare *SDO* images of plages and network concentrations with *Hinode* magnetograms, which have a spatial resolution of ∼0″.3 and a noise level of order 5 G. Comparisons should also be made with *Interface Region Imaging Spectrograph* observations of the underlying chromospheric fine structure. In the near future, we anticipate that the Daniel K. Inouye Solar Telescope's visible and near-infrared spectropolarimeters will be able to measure plage magnetic fields with unprecedented sensitivity and with spatial resolution down to ≲0″.1.

We are indebted to the referee for comments and to J. T. Hoeksema, Y. Liu, and V. Martínez Pillet for helpful correspondence regarding magnetograph measurements. This work was supported by the NASA H-GCR program and by the Chief of Naval Research.

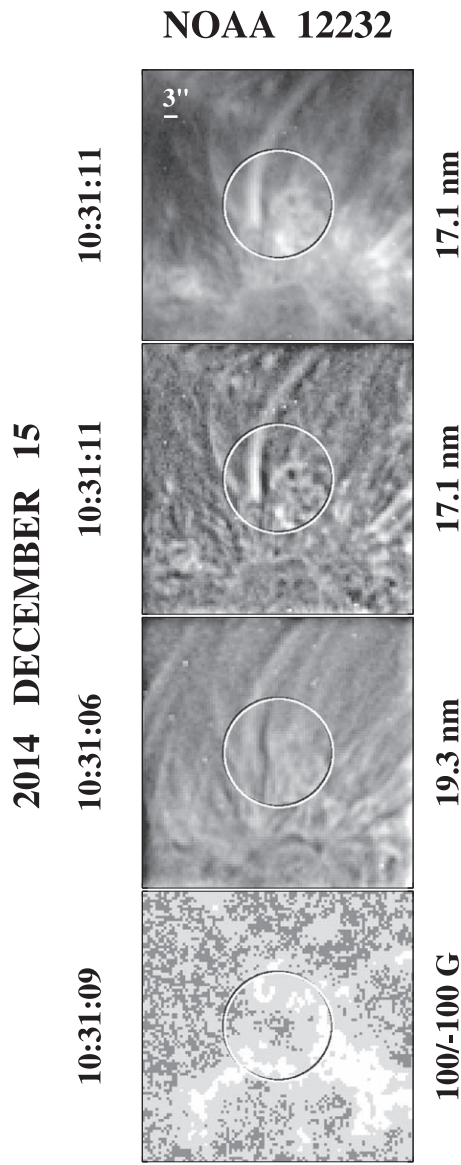

**Figure 11.** Bright moss overlying weak, mixed-polarity field inside NOAA 12232, 2014 December 15 (see also the top left corner of the images of Figure 6). The circled area has radius 12″ and is centered at (235″, 235″). Top panel: Fe IX 17.1 nm image recorded at 10:31:11 UTC. Second panel: sharpened version of the same image. Third panel: sharpened Fe XII 19.3 nm image recorded at 10:31:06 UTC. Bottom panel: HMI magnetogram recorded at 10:31:09 UTC and saturated at ±100 G.

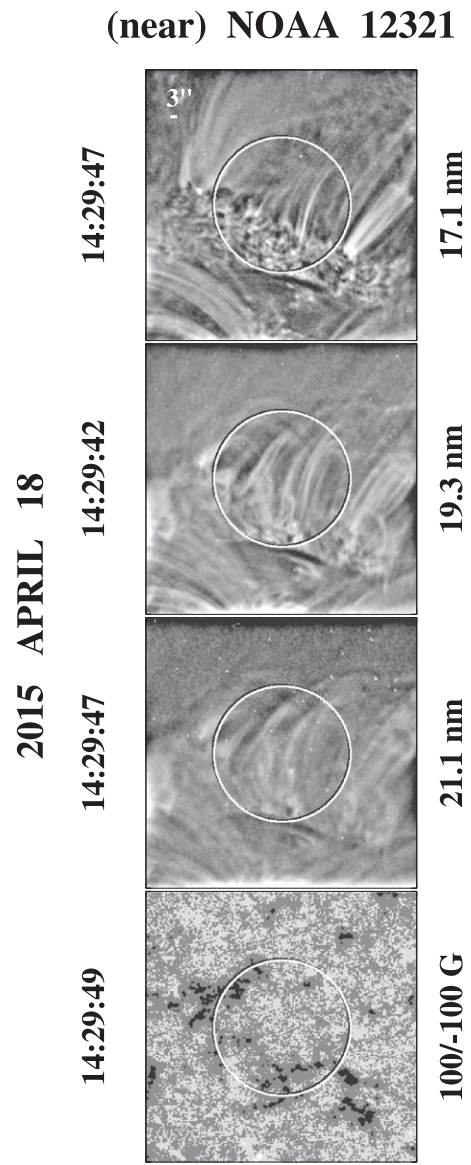

**Figure 12.** Moss-like region associated with supergranular magnetic network located just north of NOAA 12321, 2015 April 18. The circled area has radius 30″ and is centered at (265″, 391″). Top panel: sharpened 17.1 nm image recorded at 14:29:47 UTC. Second panel: sharpened 19.3 nm image recorded at 14:29:42 UTC. Third panel: sharpened 21.1 nm image taken at 14:29:47 UTC. Bottom panel: HMI magnetogram recorded at 14:29:49 UTC and saturated at ±100 G. In this case, the looplike fine structure seen inside the moss is consistent with the presence of positive-polarity flux elements scattered in and around the negative-polarity network.

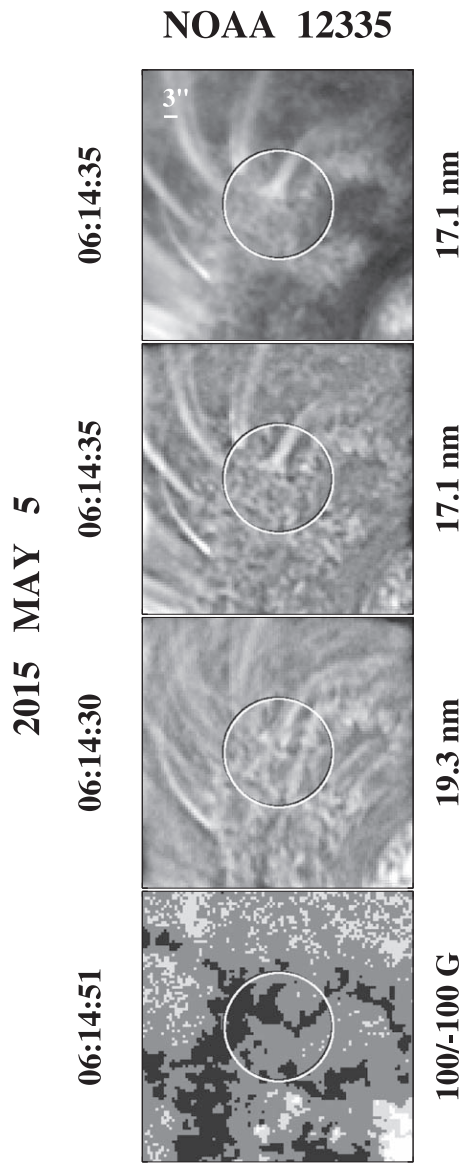

**Figure 13.** Moss region inside NOAA 12335, 2015 May 5. The circled area has radius 12″ and is centered at (−376″, −152″). Top panel: 17.1 nm image taken at 06:14:35 UTC. Second panel: sharpened version of the same image. Third panel: sharpened 19.3 nm image taken at 06:14:30 UTC. Bottom panel: HMI magnetogram recorded at 06:14:51 and saturated at ±100 G. The inverted-Y structure is located in strong negative-polarity plage, but the underlying moss extends southward into areas of weaker photospheric field.

**ORCID iDs**

Y.-M. Wang https://orcid.org/0000-0002-3527-5958